\documentclass[a4paper,fleqn,usenatbib]{mnras}
\usepackage{newtxtext,newtxmath}
\usepackage[T1]{fontenc}
\usepackage{ae,aecompl}

\usepackage{graphicx}	
\usepackage{amsmath}	
\newcommand{\xten}[1]{\mbox{$\times 10^{#1}$}}

\newcommand{\ltappeq}{\raisebox{-0.6ex}{$\,\stackrel
{\raisebox{-.2ex}{$\textstyle <$}}{\sim}\,$}}
\newcommand{\gtappeq}{\raisebox{-0.6ex}{$\,\stackrel
{\raisebox{-.2ex}{$\textstyle >$}}{\sim}\,$}}

\title[Simplified chemistries for H$_2$O and CO]
{Equilibrium simplified chemistries for H$_2$O and CO in three-phase 
astrochemical models}

\author[J.M.C. Rawlings, E. Keto and P. Caselli]
{J.M.C.~Rawlings,$^{1}$\thanks{E-mail: jcr@star.ucl.ac.uk}
E.~Keto,$^2$ and P.~Caselli$^3$\\
\\
$^{1}$Department of Physics and Astronomy, University College London,
Gower Street, London, WC1E 6BT, UK\\
$^2$Harvard-Smithsonian Center for Astrophysics, 160 Garden St., Cambridge,
MA 02420, USA\\
$^3$Max-Planck-Institut f\"{u}r extraterrestrische Physik, P.O. Box 1312, 
D-85741 Garching, Germany}

\date{Accepted 2025 August 20. Received 2025 August 9; in original form 
2025 April 6}

\pubyear{2025}

\begin{document}
\label{firstpage}
\pagerange{\pageref{firstpage}--\pageref{lastpage}}
\maketitle


\begin{abstract}
Astrochemical models can be greatly simplified, with obvious 
computational advantages, if the reaction networks for key species can be 
reduced to a bare minimum. In addition, if chemical equilibrium holds, then 
simple analytical solutions can be formulated. These have particular advantages
in the application to complex models evolving over multi-point spatial grids.
In this study,
the equilibrium solutions to highly simplified chemical networks for CO and
H$_2$O have been re-assessed with particular attention to the formulation
of the ice desorption rates in the context of `three-phase' gas-grain 
astrochemical models.
The analytical solutions have also been updated to 
account for the chemically inert reservoir of molecules below the surface
ice layers, and to include the effects of reactive desorption. 
We find that a very close match is obtained to the results from
detailed three-phase models of the time-dependent astrochemistry,
and the abundances are typically accurate to within a factor of two over 
the entire range of densities and extinction that are applicable to dense 
clouds and young star-forming regions. 
In addition, these solutions give accurate results over most of the range 
of conditions even for systems undergoing rapid dynamical evolution.
Although there are some caveats of applicability, we therefore recommend 
that these solutions be used in models of cold molecular environments 
where the rapid calculation of the abundances of CO, H$_2$O and atomic coolants
is helpful.
\end{abstract}


\section{Introduction}
\label{sec:intro}

The temporal and spatial evolution of chemical abundances in dynamically evolving 
astrophysical sources are, in general, defined by large, interconnected 
networks of chemical reactions that are described by complex, non-linear 
differential equations.
However, these formulations can be demanding of computational 
resources, and it is often only necessary to know the abundances of several
dominant species, such as H$_2$O and CO, if we wish to understand the chemistry 
of other carbon- and oxygen-bearing molecules.
In addition, CO rotational line cooling is important in the molecular gas, 
whilst [CII] and [OI] fine structure lines are the main coolants for the atomic 
gas.
Moreover, for some chemical species the chemistries are, to a certain 
degree, separable and may also be described adequately by a much simplified 
network of reactions.
In addition, if the condition of chemical equilibrium can be assumed, then 
simple analytical solutions can be derived.
The fast calculation of the abundances of key species
is particularly beneficial when the chemistry is coupled to complex numerical 
models (e.g. of protostellar and protoplanetary cloud evolution).
The computational demands of multi-point hydrodynamical/chemical models can 
be significantly reduced if it is possible to define these simplified
networks and associated analytical solutions. These definitions also show 
the inner workings and essential chemistry in chemical models with large 
numbers of species, 
giving insight into the main formation and destruction processes.

This was investigated and applied in a series of papers \citep{KC08,KC10,KRC14,RKC24}, 
in which simplified networks were proposed and applied for the key molecular 
species CO and H$_2$O.
In these cases, the simplified networks were used to describe the behaviours 
of the abundances of gas-phase and solid-state CO and H$_2$O, as well as C, 
C$^+$, O and OH.
In \citet{KC08} a simplified network was constructed for the CO chemistry and
an analytical solution was derived, on the assumption of chemical equilibrium.
\citet{KRC14}, hereafter KRC14, determined a simplified network 
and analytical (equilibrium) solution for four species (H$_2$O, 
H$_2$O$_{\rm s.}$, O and OH) in 10 reactions. 
The chemistry consisted of freeze-out reactions, followed by rapid hydrogenation, 
desorption and photodissociation, augmented by two high temperature 
neutral-neutral reactions. The analytical solutions were successfully compared 
against the full models of \citet{Holl09}.
These studies presented a novel and effective treatment of the previously 
intractable gas-grain chemistry, so that the modelling correctly predicted the 
strengths and shapes of the H$_2$O emission line profile in the pre-stellar core 
L1544 before it was actually observed \citep{Cas10,Cas12}.

For the chemistry of H$_2$O, the use and applicability of reduced networks 
was described in the review paper of \citet{vD21}.
In the cold, dense environments of young prestellar and protostellar cores,
H$_2$O ice (H$_2$O$_{\rm s.}$) is mainly formed in-situ on dust grains, following 
the accretion of O atoms (and OH radicals). This typically dominates over
low temperature ion-neutral and high temperature neutral-neutral gas phase
reactions \citep{vD13}.

\citet{Schm14} used an even more simplified network of just three species 
(H$_2$O, O and H$_2$O$_{\rm s.}$) and four reactions in their {\sc SWaN} 
(Simplified Water Network) model to describe the H$_2$O and H$_2$O$_{\rm s.}$ 
abundances (only). This particular model was designed for the study of 
the cool, outer envelopes of low-mass (Class 0 and I) protostellar 
cores that contain a central heat source. 
In these environments, the surface formation of H$_2$O$_{\rm s.}$ is 
prohibited in the warmer parts of the envelope (T$\gtappeq$15K) due to the 
low binding energy of oxygen atoms (taken to be 800K), whilst thermal 
desorption of H$_2$O$_{\rm s.}$ is highly efficient in the inner regions 
(where T$\gtappeq 100$K). This model was found to give reasonably accurate
results, although some significant discrepancies with the results 
obtained from multi-layer three-phase models were noted \citep{vD21}.

In \citet{RKC24}, hereafter RKC24, we showed that (subject to the corrections 
described below), replacing more complete chemistries by 
the simplified H$_2$O chemistry, as defined in KRC14 and an 
augmented form of the simplified CO chemistry as defined in \citet{KC08}
yielded abundance profiles that closely matched those obtained 
using a chemical model with a detailed gas-grain chemistry 
({\sc STARCHEM}, discussed in section~\ref{sec:compare} below) coupled to 
a hydrodynamic model for L1544 \citep[described in][]{KC10}.  
This was shown to hold over a very wide range of densities and extinctions.

In that study, we also showed that for isolated pre-stellar cores,
such as L1544, the chemistry of most small molecular species can be treated as 
being in quasi-equilibrium - that is to say adjusting to the instantaneous 
physical conditions, with little dependence on the historical evolution.
This justifies the assumption of chemical equilibrium made in \citet{KC08} and
KRC14 and the evaluation of the abundances as solutions to
simple algebraic expressions. These allow the rapid calculation of the abundances 
as functions of the rate coefficients and the density, temperature and extinction.
It also allows us to establish the thermal balance and gas temperature, as 
was used in the modelling of the H$_2$O line profiles and strengths in L1544 (KRC14).

The accuracy of the analytical solutions was also tested in RKC24 
(see Figure 5) and was found to work quite well. However, some significant 
discrepancies with respect to the results obtained with {\sc STARCHEM} were 
noted; particularly for gas-phase CO and H$_2$O which are over-estimated in the 
regions where thick ice mantles are present.
This is primarily due to the different representations of the gas-grain 
interactions. We address these differences in this study.

In this paper we re-formulate the equations for the condition of chemical 
equilibrium, paying particular attention to the desorption terms in the 
`three-phase' representation of gas-grain interactions, where the three 
phases are the gas, the active (surface) ice layers and the inert 
sub-surface layers. 
There have been several developments and variants of the earliest 
three-phase models of \citet{HH93} that have, for example, included the
effects of bulk ice mantle chemistry \citep[e.g.][]{Garr13,RWH16}.
Recent studies can be broadly divided into `bulk' three-phase models that
group multiple ice layers as a single `phase' \citep{FDV17} and `multilayer'
three-phase models that consider the composition of each individual ice layer
within the accreted mantle \citep[][RKC24]{TCK12}. The models also include 
the possiblilty of diffusion between ice layers and diffusive surface 
chemistry.
Subsequent studies have made further refinements, such as the inclusion of 
non-diffusive chemistry in the surface layers \citep[e.g.][]{Shin18}.

Here, we present 
the form that is applicable to regions where the dust grains have accreted 
less than one monolayer of ice. We then specify the necessary modifications 
that are necessary to make the model applicable to regions where thicker ice 
mantles are present, within the three-phase paradigm, but without 
having to apply the complexity of those models. We also consider the 
additional effects of reactive desorption. We compare the results from 
these equilibrium analytical solutions to more complex chemical models 
({\sc STARCHEM}) and consider the viability of using these solutions to 
dynamically evolving systems.

The structure of the paper is as follows: the simplified networks for the H$_2$O and 
CO chemistries are described in section \ref{sec:networks}, and the equilibrium
solutions are discussed in \ref{sec:equilib}, together with the required corrections
for application to the partial ice monolayer regime. 
Section \ref{sec:compare} compares the results to those obtained with the 
{\sc STARCHEM} chemical model and specifies a solution for icy regions. 
The effects of including reactive desorption are shown in
section \ref{sec:rdes} and the application to dynamically evolving regions is 
discussed in section \ref{sec:dynamics}. Section \ref{sec:limits} outlines the 
various limitations and assumptions in the model, and our conclusions are given in 
section \ref{sec:summary}.

\section{The simplified network chemistry}
\label{sec:networks}

In KRC14 we proposed a simplified chemical network for oxygen in cold
molecular environments, where gas phase reactions are negligible
in comparison to grain surface hydrogenation. In this model it is assumed that 
all oxygen atoms and OH radicals are instantly converted to H$_2$O on the surface 
of dust grains. 
This network, coupled to a radiative transfer model, was successfully applied 
to explain observations of the H$_2$O line profile and strength towards the 
well-studied pre-stellar core, L1544.

The simplified reaction network for H$_2$O and associated rate co-efficients are,

\noindent
\begin{tabular}{ll}
~\\
${\rm O + H_2 \to OH + H}$ & $k_1 ~({\rm cm}^3{\rm s}^{-1})$ \\
${\rm OH + H_2 \to H_2O + H}$ & $k_2 ~({\rm cm}^3{\rm s}^{-1})$ \\
${\rm OH + h\nu \to O + H}$ & $P_{\rm OH} ~({\rm s}^{-1})$ \\
${\rm H_2O + h\nu \to OH + H}$ & $P_{\rm H_2O} ~({\rm s}^{-1})$ \\
${\rm O + grain \to H_2O_{(s.)}}$ & $F_{\rm O} ~({\rm cm}^3{\rm s}^{-1})$ \\
${\rm OH + grain \to H_2O_{(s.)}}$ & $F_{\rm OH} ~({\rm cm}^3{\rm s}^{-1})$ \\
${\rm H_2O + grain \to H_2O_{(s.)}}$ & $F_{\rm H_2O} ~({\rm cm}^3{\rm s}^{-1})$ \\
${\rm H_2O_{(s.)} \to H_2O}$ & $D_{\rm H_2O} ~({\rm s}^{-1})$ \\
${\rm H_2O_{(s.)} \to OH}$ & $D_{\rm OH} ~({\rm s}^{-1})$ \\
${\rm H_2O_{(s.)} \to O}$ & $D_{\rm O} ~({\rm s}^{-1})$ \\
~\\
\end{tabular}

The first two reactions have activation barriers and are negligibly slow at dark 
cloud temperatures, but are included to provide detailed balance back reactions 
for the photodissociation reactions, as well as allowing applicability to warmer
environments.
As discussed in KRC14 the desorption of H$_2$O to yield O (with a rate
equal to that for the desorption to yield H$_2$O) is an arbitrary
addition, to provide a back reaction for the freeze-out of atomic oxygen.

Similarly, a simplified chemical network describing the chemistry of CO was defined 
in \citet{KC08}, and used to model the CO line profiles in L1544 in \citet{KC10}. 
The network was updated and expanded in RKC24. 
The (revised) reaction scheme that we proposed for the CO chemistry
is:

\noindent
\begin{tabular}{ll}
${\rm C + h\nu \to C^+ + e^-}$ & $P_{\rm C} ~({\rm cm}^3{\rm s}^{-1})$ \\
${\rm C^+ + e^- \to C + h\nu}$ & $k_3 ~({\rm cm}^3{\rm s}^{-1})$ \\
${\rm C + OH \to CO + H}$ & $k_4 ~({\rm cm}^3{\rm s}^{-1})$ \\
${\rm C^+ + OH \to CO^+\dots\to CO}$ & $k_5 ~({\rm cm}^3{\rm s}^{-1})$ \\
${\rm CO + h\nu \to C + O}$ & $P_{\rm CO} ~({\rm s}^{-1})$ \\
${\rm CO + grain \to CO_{(s.)} }$ & $F_{\rm CO} ~({\rm cm}^3{\rm s}^{-1})$ \\
${\rm CO_{(s.)} \to CO}$ & $D_{\rm CO} ~({\rm s}^{-1})$ \\
\end{tabular}
~\\

In both these networks we denote $k_{\rm i},P_{\rm i},F_{\rm i},D_{\rm i}$ as the 
rate coefficients for gas phase two-body reactions, photoreactions, freeze-out 
and desorption, respectively. 

The photoreaction rates depend on the local flux of UV photons, denoted by 
h$\nu$ above. In the photon-dominated region, the UV flux originates outside the
clouds from the diffuse Galactic starlight and the
mutual C/CO/H$_2$ shielding factors should be taken into account.
In darker regions (typically $A_{\rm v}\gtappeq 4$ mag.), cosmic-ray 
induced processes are the dominant source of UV.
The values that we use for these rate coefficients in this study, including the
freeze-out rates and components of the desorption processes, are described in
Appendix~\ref{sec:fodes}. 

\section{The equilibrium equations}
\label{sec:equilib}

In general, the rate of change of the fractional abundance of a chemical species 
(s$^{-1}$) due to formation by chemical reactions can be described by differential
equations of the form,
\begin{equation}
\frac{dX_i}{dt} = n\sum_j\sum_k R_{i,j,k} X_j X_k + \sum_l R_{i,l} X_l 
+ D_i f_i 
\label{eqn:master}
\end{equation}
with similar expressions for the destruction rates.

In this equation the first term corresponds to two-body processes (e.g. bimolecular
reactions and freeze-out) where $n$ is the (hydrogen nucleon) density, the second 
term corresponds to processes where the rate only depends on the abundance of one 
species (e.g. photoreactions and cosmic-ray ionization reactions), and the final 
term corresponds to ice desorption. Note that for the desorption processes, 
the rates depend on the surface coverage of the dust grains by the relevant 
ice species ($f_{\rm i}$), rather than the solid-state abundance ($X_{\rm i}$). 
%
%
$X_{\rm i}$ is the fractional abundance of species $i$, relative to hydrogen 
nucleons, and
the units of $R_{i,j,k}$, $R_{i,l}$ and $D_{\rm i}$ are cm$^3$s$^{-1}$, s$^{-1}$ and 
s$^{-1}$, respectively.

In equilibrium, the condition of detailed balance implies that the net time 
derivatives are zero for all species ($dX_{\rm i}/dt=0$) resulting in a set
of N algebraic equations for N species. As there is also a conservation equation
(for the total oxygen, or carbon abundance), the system is 
over-determined and only N-1 of the 
algebraic equations are required to solve for the variables. For the oxygen 
network these are the abundances of O, OH, H$_2$O and H$_2$O$_{\rm (s.)}$,
whilst for the carbon network we solve for C, C$^+$, CO and CO$_{\rm (s.)}$.

\subsection{Algebraic solutions}
\label{sec:update}

The original formulation of the detailed balance solution to the simplified network 
of oxygen chemistry reactions was presented in Appendix B of KRC14 and
followed an approach similar to `two-phase' models where there is no 
distinction between the chemically active surface layers of the ice mantles 
and the chemically inert sub-surface layers.
These previous studies were groundbreaking in that they showed the viability of 
using equilibrium solutions of the simplified network chemistries for both 
H$_2$O and CO \citep[e.g.][]{Brod07}. 
Subsequent studies then found that the abundance profiles, in combination 
with radiative transfer modelling, could accurately predict spectral line 
emissions from dense molecular cloud cores \citep{KC08,KC10,Cas12}. 

However, whilst the rates for the gas-phase and freeze-out reactions are all
proportional to the abundances, the desorption processes (the last term in 
equation~\ref{eqn:master}) are more complicated, in that the rate depends on
the fractional surface coverage of a species on the dust grains ($f_{\rm i}$) 
which is not necessarily proportional to the total abundance of a solid-state 
species.
With the two-phase assumption, it is reasonably accurate to numerically equate 
the proportion of the grain surface that is covered by a species to the 
fraction of the ice mantle that is composed of that species: 
\begin{equation}
f_{\rm i} = \left[ \frac{X_{\rm i}}{X_{\rm ice,total}} \right],
\label{eqn:fi}
\end{equation}
where $X_{\rm ice,total}$ is the total fractional abundance of solid-state 
species and, for the dominant ice mantle components, $f_{\rm i}$ is of 
order unity \citep[e.g.][]{RRVW07}.
However, this assumes that there is complete coverage of the dust grains 
by at least one monolayer of ice.
\footnote{There is a potential source of misunderstanding here:
$f_{\rm i}$ is formally defined as the `fraction of the surface that is 
covered by species $i$'.
In the two-phase approximation it is usually assumed that $f_{\rm i}$ 
is numerically equal to the `fraction of the ice mantle that is composed of 
species $i$'.
This is actually {\bf only} true in the saturation limit of $\sum_i f_{\rm i}=1$.
In the partial monolayer regime $f_{\rm i}$ may be very much less than 
this value.}
With this assumption. the desorption rate can then be expressed as a function of the 
solid-state abundance of a species, although there is still a non-trivial 
complication in that the total ice abundance is dependent on the abundances of
other species.


In the so-called `three-phase' model of gas-grain interactions, the chemically 
active surface ice layers and the inert inner ice mantle are chemically 
disconnected.
The desorption reactions are predominantly surface processes so that,
once a few layers of ice have accreted, the desorption rates
saturate ($\sum_i f_i =1$) and are {\em not} dependent on the total solid-state 
abundance of a species ($X_{\rm i}$). 
Rather they are defined by the abundance in the surface layers only, i.e. the 
fraction of the grain surface that is covered by that species ($f_{\rm i}$). 
The sub-surface layers are chemically inaccessible, so that the total fractional
abundance of a solid-state species ($X_{\rm i,s}$) does not enter into the rate 
equations and is not constrained by the equilibrium solution to those equations.

\subsection{Partial monolayer solution}
\label{sec:mono}

Although equation (\ref{eqn:fi}) and the approximation that the desorption 
rates have a linear dependence on total ice abundances ($X_{\rm i}$) is not 
generally applicable, there is a special situation when dust grains 
are covered by less than one layer of ice.
In these circumstances the fraction of the adsorption sites on the surface 
of the dust grains that are occupied by that species is, indeed, proportional to 
the total solid-state abundance of that species in which case the equation
is valid, provided the appropriate normalisation is made.

If $N_{\rm s}$ is the surface density (cm$^{-2}$) of adsorption binding sites 
then $\sigma_{\rm A}N_{\rm s}$ (where $\sigma_{\rm A}$ is the dust grain surface 
area per hydrogen nucleon) gives the number of binding sites per hydrogen nucleon 
- effectively the `fractional abundance of surface binding sites'.
So long as the total ice abundance is less than this value, we may assume that 
there is less than one monolayer of ice on the grains.

In the partial monolayer regime the fractional coverage of the surface layer 
by species $i$ is then given by
\begin{equation}
f_{\rm i} = X_{\rm i,s}/(\sigma_{\rm A}N_{\rm s}), {\rm ~~subject~to~~} 
\sum_i f_i \leq 1,
\label{eqn:cover}
\end{equation}
where $X_{\rm i,s}$ represents the fractional abundance of species $i$, apportioned
to the surface layer, and $\sigma_{\rm A}N_{\rm s} \sim 2\times 10^5$.
In this study we limit the ice composition to just two species; H$_2$O$_{\rm s.}$
and CO$_{\rm s.}$.
%

Equation~(\ref{eqn:cover}) assumes that; (i) that the ice accumulates in 
concentric layers, and (ii) that the H$_2$O and CO ices are spatially
distinct on the surface of the grains and do not overlap in the partial
monolayer regime. These simplifications are also common to most `three-phase' 
gas-grain models. 

The advantage of these approximations is that the detailed 
balance equations for the oxygen/H$_2$O chemistry then have linear dependencies 
on the abundances of the four chemical species and can thus be expressed in 
matrix form and solved by matrix inversion. As explained above, the addition 
of the conservation equation means that one of these equations is not needed, 
so that the network can be reduced to a 4$\times$4 matrix. This is shown in 
Figure~\ref{fig:matrix}, where we have chosen to omit the equation for 
H$_2$O$_{\rm s.}$. The last row is the conservation equation. 
$X_{\rm O,total}$ is the total oxygen abundance.
The matrix also includes some expansions and clarification of the terms that
were originally used in KRC14.

\begin{figure*}
\begin{center}
\begin{equation}
\begin{pmatrix}
-(F_{\rm O}n+k_1n/2) & P_{\rm OH} & 0 & D_{\rm O}/(\sigma_{\rm A}N_{\rm s}) \\
k_1n/2 & -(F_{\rm OH}n+P_{\rm OH}+k_2n/2) & P_{\rm H_2O} & 
D_{\rm OH}/(\sigma_{\rm A}N_{\rm s}) \\
0 & k_2n/2 & -(F_{\rm H_2O}n+ P_{\rm H_2O}) & 
D_{\rm H_2O}/(\sigma_{\rm A}N_{\rm s}) \\
1 & 1 & 1 & 1
\end{pmatrix}
\begin{pmatrix}
X_{\rm O} \\
X_{\rm OH} \\
X_{\rm H_2O} \\
X_{\rm H_2O_{(s.)}}
\end{pmatrix}
=
\begin{pmatrix}
0 \\
0 \\
0 \\
X_{\rm O,total}
\end{pmatrix}
\end{equation}

\begin{align}
{\rm X_C + X_{C^+} + X_{CO} + X_{CO_{(s.)}}} & = X_{\rm C,total}\\
{\rm k_3 n X_{C+}^2 + P_{CO}X_{CO} -(P_C + k_4 X_{OH}n)X_C} & = 0 \\
{\rm P_C X_C -k_3 nX_{C^+}^2 - k_5 X_{OH}n X_C^+} & = 0 \\
{\rm F_{CO}nX_{CO} - D_{CO}X_{CO_{(s.)}}/(\sigma_{\rm A}N_{\rm s})} & = 0 
\end{align}

\caption{Matrix representation of the simplified network equilibrium solution
for the H$_2$O chemistry, and equilibrium equations for the CO chemistry.}
\label{fig:matrix}
\end{center}
\end{figure*}

The elements of the matrix of rate coefficients all have units of $s^{-1}$.
The factors of $1/2$ in some of the terms originates
from the assumption that the gas is fully molecular, so that $n(H_2)\sim 0.5n$.
%
The solution can then be obtained by inverting the 4$\times$4 matrix, with
the abundances ($X_{\rm i}$) given by the final column of the matrix inversion.

In the carbon network described above, the abundance of electrons ($X_{\rm e^-}$) 
is not calculated explicitly and, following RKC24, we make the assumption 
that it is equal to that for C$^+$ (i.e. $X_{\rm e^-}=X_{\rm C^+}$),  
consistent with the findings of \citet{GC12}.
However, the addition of a non-linear dependence (the rate for the second 
reaction being proportional to $X_{\rm C^+}^2$) means that the equilibrium
equations cannot be expressed in simple matrix form, but instead must be 
solved as a quadratic.

The equations for carbon conservation, C, C$^+$ and CO$_{\rm s.}$ are 
also given in Figure~\ref{fig:matrix},
where $X_{\rm C,total}$ is the total elemental abundance of carbon.
As with the H$_2$O network, a fifth reaction (here for CO formation/destruction) 
could also be specified but, as we already have four equations in four unknowns, 
it would be redundant.
%
%
The algebraic expansions for both solutions are given in 
Appendix~\ref{sec:analytic}.

The abundance of OH used in these expressions is first of all determined from the 
solution for the H$_2$O chemistry described above and is treated as being 
independent of the carbon chemistry. This approximation is satisfactory, and the 
OH abundance will not be strongly affected by the reactions with C or C$^+$, so 
long as $X_{\rm O,total}>X_{\rm C,total}$. 

This approach was applied for the results that were shown in RKC24 and 
much better fits to the numerical results of {\sc STARCHEM} were obtained,
except in the regions where substantial freeze-out has occured, which we discuss 
in the following section.

\subsection{Corrections and application to icy regions}
\label{sec:ices}

The formulation described above will hold so long as there is less than one 
monolayer of ice covering the dust grains, otherwise the desorption rates 
will be over-estimated.
There are several modifications and corrections that can be applied, so as to 
get a closer match to the full three-phase models.

Firstly, although the CO and H$_2$O chemistries are essentially independent,
they are (weakly) linked through the reactions with OH and, in addition
the depletion of oxygen into highly stable CO (gas-phase and 
ice) will reduce the available oxygen budget for the H$_2$O chemistry.
Provided we can make the assumption that $X_{\rm o,total}>X_{\rm C,total}$ 
this can easily be accounted for by reducing $X_{\rm o,total}$ in the H$_2$O
calculations, accordingly.

Secondly, as explained above, in regions where the dust grains accumulate 
more than one ice monolayer,
the assumption that $f_{\rm i}\propto X_{\rm i}$ ceases to be applicable 
as, in the three-phase model of gas-grain interactions, the sub-surface 
layers are chemically inaccessible. In these circumstances, the desorption
rates saturate ($\sum f_{\rm i}=1$) and the formula results in their 
over-estimation by a factor that is approximately equal the ratio of the 
total number of molecules in the ice mantle to the number in the chemically 
active surface layers that can be desorbed.

The most accurate way of modelling the saturation limit would be to fix the 
desorption rates at their limit (saturation) values, which are independent 
of the ice abundances, and then solve a 3$\times$3 matrix and three 
equations for the oxygen chemistry and the carbon chemistry respectively.
This would yield the abundances of the gas-phase species (O, OH, H$_2$O, C, 
C$^+$ and CO). As the desorption rates are independent of the ice abundances the 
condition of detailed balance implies that the gas-phase abundances are
independent of the total elemental abundances ($X_{\rm i,tot}$). The ice species 
are the dominant reservoirs of oxygen and carbon in these conditions, so that 
$X(H_2O_{\rm (s.)})$ and $X(CO_{\rm (s.)})$ would then be determined from 
the two elemental conservation equations.

Instead, to simplify the procedure, we propose an alternative solution of 
re-scaling the desorption rates in the existing formulation.
We find that, in all cases, this results in the accurate determination of 
both the gas-phase and the solid-state abundances, with near-identical results
obtained with the two methods.

Thirdly, 
to determine the necessary scaling factors we may also need to compensate for the 
increase in effective grain size, if the thickness of the ice mantle is comparable 
or greater than the radius of the bare grains. This is applicable to small
grains ($a\ltappeq 0.01\mu$m), as modelled this study, and is discussed in 
Appendix~\ref{sec:growth}.

\subsubsection{The procedure}
\label{sec:proc}

For the reasons described above and recognising that the cross terms in 
the chemistry mean that the H$_2$O and CO chemistries are not completely
independent, we therefore adopt a stepwise approach:
After obtaining the solution to the equations described above, the procedure 
is to repeat the calculation with the following steps: 
\begin{enumerate}
\item Offset the oxygen abundance in the matrix solution:
$X_{\rm O,total} = X_{\rm O,total,0} - X_{\rm CO} - X_{\rm CO(s.)}$
where $X_{\rm O,total,0}$ refers to the initial, undepleted, oxygen abundance,
\item Calculate the total ice abundance: $X_{\rm ice} = X_{\rm H_2O_{(s.)}} + 
X_{\rm CO_{(s.)}}$,
\item Determine whether there is greater than one monolayer of ice
(i.e. if $X_{\rm ice}>\sigma_{\rm A,0}N_{\rm s}$),
\item If so, scale the desorption rates by a scaling factor $R_{\rm scale}$, and
\item Re-calculate the analytical solutions.
\end{enumerate}

The scaling factor for the desorption rates is given by
\begin{equation}
R_{\rm scale} = \left( n_{\ell}\frac{N_{\rm s}\sigma_{\rm A}}
{X_{\rm ice}}\right) \left( \frac{X_{\rm i}}{X_{\rm ice}}\right),
\end{equation}
where the first factor (which is obviously limited to being $\leq 1$) specifies 
the (chemically active) surface to volume ratio and the second
factor gives the fraction of the ice that is composed of species $i$. 
In this expression, $\sigma_{\rm A}$ has a dependence on $X_{\rm ice}$,
as given by equations (\ref{eqn:growth}) and (\ref{eqn:area}) in 
Appendix~\ref{sec:growth}. This is included for the case of a small
mean grain radius ($a\ltappeq 0.01\mu$m), as in this study, but can be 
ignored for larger grains ($a\sim 0.1\mu$m) in which case 
$\sigma_{\rm A}\sim \sigma_{\rm A,0}$. 
$n_{\ell}$ is the number of surface layers that are 
considered to be chemically active, and prone to desorption (typically 
$n_{\ell}=1-3$).
A further refinement, which is not investigated in this study, would be to 
sub-divide the desorption rate and apply a value of $n_{\ell}$ that is 
applicable for each desorption mechanism.

Here we have made several assumptions:
\begin{enumerate}
\item[(a)] that the composition of the surface ice layer is the same as the bulk
composition of the ice mantle,
\item[(b)] that the ices form in uniform concentric layers of equal thickness 
($\Delta r$), 
\item[(c)] that the ices are predominantly composed of H$_2$O and CO, and
\item[(d)] that $X_{\rm O,total}>X_{\rm C,total}$.
\end{enumerate}

Obviously, the abundances determined in the second calculation will result in
modifications to the oxygen abundance offset (step (i) above) and the values 
of $X_{\rm ice}$ (and possibly $\sigma_{\rm A}$) in $R_{\rm scale}$, so this 
should be considered the first iteration.
However, we find that subsequent iterations result in negligibly small 
variations in all of the cases that we have considered in this study, and so are
not necessary.

\section{Comparison with detailed models}
\label{sec:compare}

To assess the accuracy of the equilibrium solution to the simplified chemistry
network we compare our results with those obtained from a full time-dependent
chemical model that incorporates three-phase gas-grain interactions.
This method is described in RKC24 and employs the {\sc STARCHEM} code to 
model the chemical evolution of the quasi-statically contracting starless 
core L1544.

In this model {\sc STARCHEM} is configured to follow the chemical and physical
evolution in Lagrangian streamlines of 100 grid points in spherical (1D)
symmetry, for an assumed core size of $\sim$0.3pc. The hydrodynamic flow and 
physical profiles are defined by a well-established model of quasi-static 
contraction in prestellar cores \citep{KC10}. 
This model gives the radial profiles of the gas density, the temperatures 
of the gas and dust (for which we assume a single size population), the 
extinction and the infall velocity. These profiles are shown in Figure 1
of RKC24. 
Note that the dust temperature is $<10$K at all positions in this simulation
so that thermal desorption is negligible, for both CO and H$_2$O. 
The chemistry and dynamics are evolved for 0.61Myr, so as to 
match the observationally inferred conditions in L1544.
%
This model has been 
empirically constrained and validated by comparison to observations of 
L1544 (the predicted abundance profiles and line strengths of C$^{16}$O, 
C$^{17}$O and C$^{18}$O (1-0) and H$_2$O 567GHz (1$_{10}-1_{11}$); 
\citep{KC10,KRC14,KCR15}. 

The model is described in detail in RKC24, but to summarise; 
the chemistry (of 81 gas-phase and 17 solid-state species in $\sim$1250 
chemical reactions) includes a 
comprehensive description of gas-phase, freeze-out, the various desorption
processes, and a simplified form for the network of surface reactions.
The compositional structure of the (assumed spherical) dust grain ice mantles 
is also followed as a function of position and time, and we make the 
so-called `three-phase' approximation, where only the surface layer(s) of 
the ices are treated as being chemically active and connected to the 
gas-phase, whilst the sub-surface layers are considered to be chemically inert.
{\sc STARCHEM} employs a multi-layer three-phase model, that considers 
the detailed composition of individual ice layers.
For simplicty, in this particular comparison we have restricted the 
desorption to the top surface layer only ($n_{\rm l}=1$).
The outer parts of the core ($r\gtappeq 0.07$pc) are treated as a photon 
dominated region (PDR) and approximate shielding functions are included for 
the effects of H$_2$, CO and CO  self- and mutual-shielding effects for 
photodissociation and photoionization.

\begin{table*}
\caption{Values of key model parameters in the {\sc STARCHEM} model.\\
$^\dag$Note that two values are given for $\epsilon_{\rm OH}$ and
$\epsilon_{\rm H_2O}$, corresponding to $<1$ and $\geq 1$ ice monolayers.}
\begin{tabular}{|l|l|}
\hline
 Total carbon abundance ($C/H$) & $1.5\times 10^{-4}$ \\
 Total oxygen abundance ($O/H$) & $2.5\times 10^{-4}$ \\
 Initial CO$:$H ratio & $0.95X_{\rm C_{tot}}$ \\
 ISRF photon flux ($I_0$) & 1.0$\xten{8}$ photons cm$^{-2}$ s$^{-1}$ \\
 Photon flux/ISRF ($G_0$) & 1.0 \\
 Cosmic-ray induced photon flux ($I_{\rm cr}$) & 4875.0 photons cm$^{-2}$ s$^{-1}$ \\
 Binding site surface density ($N_{\rm s}$) & 1.0$\xten{15}$ cm$^{-2}$ \\
 Mean ice monolayer thickness ($\Delta r$) & 3.7\AA \\
 Number of chemically active surface layers ($n_{\rm l}$) & 1 \\
 Cosmic-ray ionization rate for H$_2$ ($\zeta_{\rm cr}$) & 1.3$\xten{-17}$ s$^{-1}$ \\
 Initial dust surface area per H nucleon ($\sigma_{\rm A,0}$) & 6.0$\xten{-21}$ cm$^2$ \\
 Initial RMS grain radius ($a_0$) & 0.01$\mu$m \\
 Fraction of adsorbed CO that is converted to CO$_2$ or CH$_3$OH & 0 \\
 Photodesorption yield for H$_2$O$_{\rm (s.)}\to$ H$_2$O ($Y_{\rm H_2O}$) & 3.33$\xten{-4}$\\
 Photodesorption yield for H$_2$O$_{\rm (s.)}\to$ OH ($Y_{\rm OH}$) & 6.67$\xten{-4}$\\
 Photodesorption yield for H$_2$O$_{\rm (s.)}\to$ O ($Y_{\rm O}$) & 3.33$\xten{-4}$\\
 Photodesorption yield for CO$_{\rm (s.)}\to$ CO ($Y_{\rm CO}$) & 1.0$\xten{-3}$\\
 $^\dag$Reactive desorption efficiency for OH ($\epsilon_{\rm OH}$) & 0.5,0.25\\
 $^\dag$Reactive desorption efficiency for H$_2$O ($\epsilon_{\rm H_2O}$) & 0.8,0.3\\
 Yield for H$_2$-formation driven desorption ($Y_{\rm H_2}$) & 0.0 \\
\hline
\end{tabular}
\label{tab:params}
\end{table*}

\begin{figure*}
\includegraphics[scale=0.65]{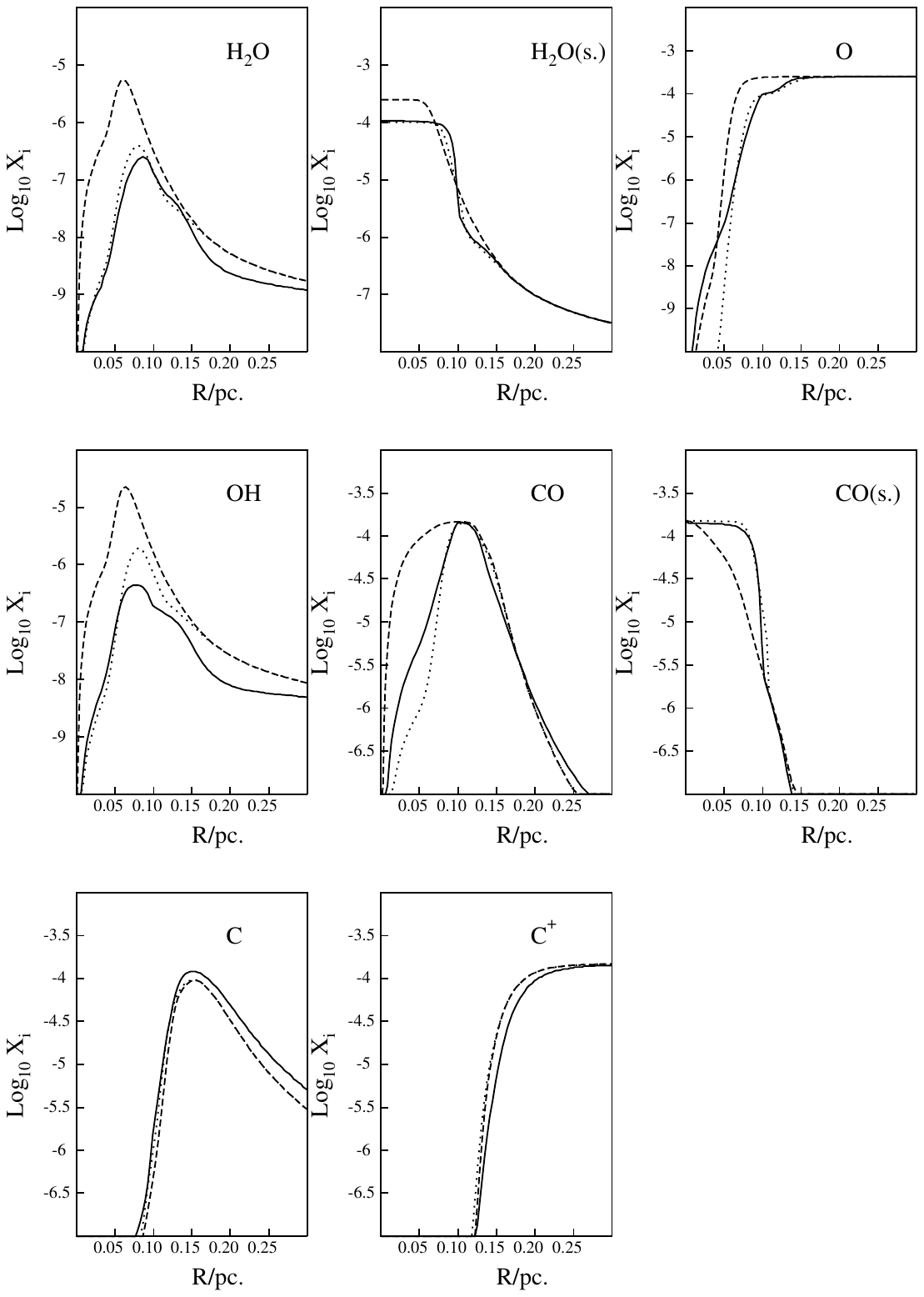} 
\caption{Comparison of equilibrium simplified networks with full chemical models. 
The solid lines show the results from the full three-phase gas-grain model 
({\sc STARCHEM}). 
The dashed lines give the results from the revised two-phase simplified network, and 
the dotted lines show the closer fit that is obtained when the algebraic solution 
includes the various three-phase modifications and corrections as discussed 
in section~\ref{sec:ices}.}
\label{fig:L1544}
\end{figure*}

We specify the model parameters that are relevant to this study in
Table~\ref{tab:params}.
Examples of the results are given in Figure~\ref{fig:L1544} which shows
the abundance profiles for all eight species (H$_2$O, H$_2$O(s.), O, 
OH, C, C$^+$, CO and CO(s.)) obtained with the {\sc STARCHEM} model 
(solid lines) together with those obtained from the solution to the 
simplified network equations for detailed balance as described above. 
For the purpose of comparison, in this model we have suppressed the 
surface chemistry of CO and other species (leading to CO$_2$ and CH$_3$OH 
formation etc.) and reactive desorption processes in {\sc STARCHEM}
(although see section~\ref{sec:rdes} below).
Results are given both before and after the corrections described in 
section \ref{sec:ices} have been applied.
In RKC24 we essentially used this formulation but, unlike the STARCHEM model,
we did not fully consider these three-phase modifications.

Using a reduced network of chemical reactions allows us to 
easily identify the dominant chemical pathways. So, for the example 
given, at the outer edge (low extinction and density), the chemistry 
is photon-dominated. The H$_2$O balance is therefore dominated by the 
freeze-out of oxygen atoms to form H$_2$O$_{\rm s.}$, which is 
photodesorbed and then photodissociated back to oxygen atoms.
For C/C$^+$, there is a photoionization/recombination balance 
(with very little CO present).
In the inner regions (high extinction and density), the freeze 
out of H$_2$O (and OH) is balanced by cosmic-ray induced photodesorption,
whilst the freeze-out of CO is balanced by cosmic-ray heating induced
desorption.

In Figure~\ref{fig:L1544} we can see that even before the modifications 
have been 
applied (dashed lines) the detailed balance formulation works extremely 
well in the outer parts of the core ($r\gtappeq$0.08pc), where there is 
less than one monolayer of ice on the dust grains. 
However in the inner, icy, regions it significantly over-estimates the 
gas-phase abundances of CO, OH and H$_2$O and under-estimates the 
solid-state abundance of CO - for the reasons given in 
section~\ref{sec:update}.

After the corrections have been applied, the figure shows 
that the three-phase model can now be well-reproduced with our 
analytical formulae; much closer fits are obtained (dotted lines) for 
all four species, as well as better fits for the O and H$_2$O(s.) profiles.
The close fit to the C and C$^+$ profiles also justifies the assumption 
made in section~\ref{sec:mono} that $n(C^+)\sim n(e^-)$.

The analytic solution still predicts slightly too much OH at the abundance 
peak near 0.06pc.
The analytical decoupling of the H$_2$O and CO chemistries means that 
solution for the H$_2$O chemistry does not take account of the reduction 
of H due to the reactions with C and C$^+$ in the CO chemistry, but the 
effects are found to be minimal. Indeed, an analysis shows that, as
OH is a reactive radical there are other loss channels for OH that 
the simplified network does not consider.
With this exception, the modified model yields abundance profiles that are very 
close (within a factor of $2\times$ at all positions) to those obtained with the 
STARCHEM model.

The remaining (small) discrepancy in the gas-phase CO profile at small radii 
(where the ice abundances are large) is primarily a consequence of the 
inaccuracy of the assumption that the composition of the surface layers of the 
ice is the same as that in the inner mantle. In these regions the composition of 
the surface layers is dominated by CO so that 
$f_{\rm CO}>X_{\rm CO}/X_{\rm ice}$.

There is, however, one important caveat; we find that the accuracy of the
abundance profile for CO ice is subject to the condition that CO is the 
dominant carbon-bearing component in the ice mantles.
This would not be the case if there was significant accretion of atomic carbon 
(and conversion to CH$_4$) in the early stages of the evolution, before CO 
formation is complete \citep{AHRC05,Holl09,RKC24}.

\begin{figure}
\includegraphics[width=0.85\columnwidth,trim=0.5in 5.4in 4.1in 1in,clip=true]
{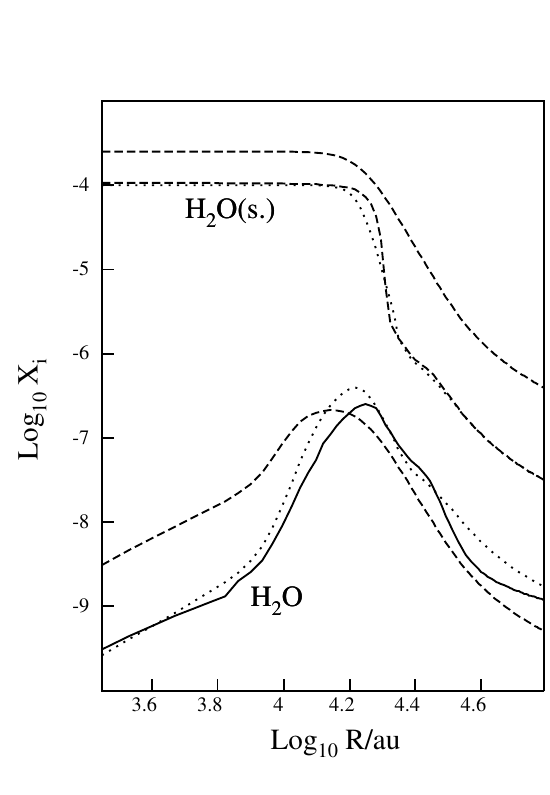}
\caption{H$_2$O and H$_2$O$_{\rm (s.)}$ profiles: from the full three-phase 
{\sc STARCHEM} chemical model (solid lines), the simplified three-phase algebraic 
solution described in this study (dotted lines), and the algebraic solution that was 
applied in KRC14 (dashed lines).}
\label{fig:krc14}
\end{figure}

For comparison, we also show in Figure~\ref{fig:krc14} the abundance profiles 
for H$_2$O and H$_2$O(s.) that are obtained with the {\sc STARCHEM} model,
together with the simplified network detailed balance solutions obtained 
(a) using the formulation that was used to generate the abundance profiles in 
KRC14, and (b) including the `three-phase' corrections, as described in this study.
This figure shows that the models actually give reasonably good matches for the
location and and strength of the H$_2$O abundance peak, but it also
highlights the difference between the `two-phase' and `three-phase' models.
Specifically, the former results in $\sim10\times$ as much H$_2$O ice in 
the outer regions, and $\sim10\times$ as much gas-phase H$_2$O
throughout most of the inner regions, as compared to the latter.
In the outermost region the difference is mostly due to there being a somewhat 
different formulation for the desorption rates in the two models, and the fact
that the KRC14 model did not take account of the oxygen abundance division
between the carbon and oxygen networks (step (i) in \S\ref{sec:proc}).
In the innermost region the difference is mainly due to the fact that the 
two-phase KRC14 model does not take account of the saturation limit for the
desorption rates.

\section{Including reactive desorption of OH and H$_2$O}
\label{sec:rdes}

The formation of chemical bonds in surface reactions is generally exothermic 
and the liberated energy (enthalpy of formation) may be sufficient to result 
in the desorption of the nascent molecule. This is probably very important for 
several molecules of astrophysical interest (most especially H$_2$CO and H$_2$O).
\citet{Min16} \citep[and also][]{Ried23} determined the enthalpies of formation 
and desorption efficiencies ($\epsilon_{\rm i}$) for the O$\to$OH and 
OH$\to$H$_2$O reactions typically to lie in the range 0.25-0.8, 
dependent on the nature of the substrate and whether or not a monolayer of ice 
has already been accumulated. 
$\epsilon_{\rm i}$ is significantly lower for reactions occurring on the surface
of ices that are predominantly composed of water ice (see Table~\ref{tab:params}).
As we assume that accreted oxygen atoms and OH radicals are instantly hydrogenated
to H$_2$O, this is something that we can therefore easily incorporate into the 
simplified network formulation by modifying the effective freeze-out rates for O and 
OH and including three extra branching reactions for the freeze-out processes:

\begin{align}
& {\rm O + grain \to OH} & & H_1 ~({\rm cm}^3{\rm s}^{-1}) \\
& {\rm O + grain \to H_2O} & & H_2 ~({\rm cm}^3{\rm s}^{-1}) \\
& {\rm OH + grain \to H_2O} & & H_3 ~({\rm cm}^3{\rm s}^{-1}) 
\end{align}

Thus, a fraction ($\epsilon_{\rm OH}$) of the adsorbed 
oxygen atoms that are hydrogenated to form OH are desorbed.
Similarly, a fraction ($\epsilon_{\rm H_2O}$) of the surface OH that is 
hydrogenated to form H$_2$O is also desorbed back into the gas-phase.

The rate coefficients for these reactions are therefore 
\begin{align}
& H_1 = \epsilon_{\rm OH}F_{\rm O} \\
& H_2 = \epsilon_{\rm H_2O}(1-\epsilon_{\rm OH})F_{\rm O} \\
& H_3 = \epsilon_{\rm H_2O}F_{\rm OH}
\end{align} 
and the net freeze-out rate coefficients for O and OH must also be reduced 
accordingly (by $H_1+H_2$ and $H_3$, respectively).

With these extra reactions and modifications, the matrix formulation is as shown 
in Figure~\ref{fig:numatrix}, where we have again chosen to omit the row 
corresponding to the detailed balance equation for H$_2$O$_{\rm s.}$.
As before, this can be inverted to find the solution, which is given in 
Appendix~\ref{sec:analytic}.

%
%

\begin{figure*}
\begin{center}
\begin{equation*}
\begin{pmatrix}
-(F_{\rm O}n+k_1n/2) & P_{\rm OH} & 0 & D_{\rm O}/(\sigma_{\rm A}N_{\rm s}) \\
k_1n/2 + H_1 n & -(F_{\rm OH}n+P_{\rm OH}+k_2n/2) & P_{\rm H_2O} & 
D_{\rm OH}/(\sigma_{\rm A}N_{\rm s})  \\
H_2 n & k_2n/2 +H_3 n & -(F_{\rm H_2O}n+ P_{\rm H_2O}) & 
D_{\rm H_2O}/(\sigma_{\rm A}N_{\rm s}) \\
1 & 1 & 1 & 1
\end{pmatrix}
\begin{pmatrix}
X_{\rm O} \\
X_{\rm OH} \\
X_{\rm H_2O} \\
X_{\rm H_2O_{(s.)}}
\end{pmatrix}
=
\begin{pmatrix}
0 \\
0 \\
0 \\
X_{\rm O,total}
\end{pmatrix}
\end{equation*}
\caption{H$_2$O chemistry matrix, with the addition of terms for reactive 
desorption.}
\label{fig:numatrix}
\end{center}
\end{figure*}

\begin{figure*}
\includegraphics[scale=0.7,trim= 0 370 0 100,clip=true]{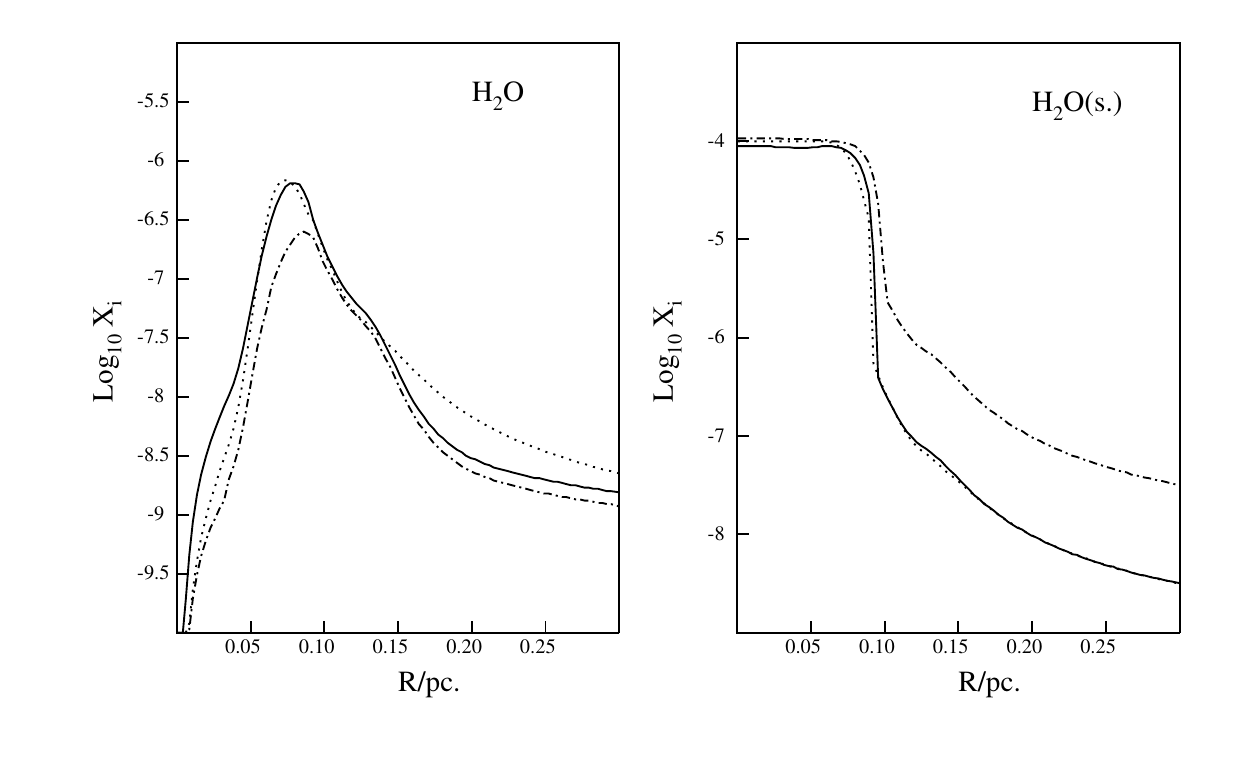} 
\caption{Results showing the effect of the inclusion of reactive desorption of 
OH and H$_2$O. 
The dotted lines show the abundances obtained from the simplified three-phase 
network model, including reactive desorption, whilst the
solid and dot-dashed lines show the results from the full STARCHEM model 
with/without the inclusion of reaction desorption, respectively.}
\label{fig:reacdes}
\end{figure*}

The effects of the inclusion of these terms is shown in Figure~\ref{fig:reacdes}, 
in which we give the abundance profiles for H$_2$O and H$_2$O(s.) as these are 
the only species that are strongly affected by the modification.
Comparing the results from the {\sc STARCHEM} model with/without reactive 
desorption (the dashed and dotted lines) shows the strong effects on the 
H$_2$O(s.) ice profile in the low extinction outer regions, and lesser effects on
the gas-phase H$_2$O profile in the dark central regions.
Comparison with the results from the simplified network model (the solid lines) 
shows that the model again gives excellent fits over the whole range of
extinction and density. 

\section{Dynamical scenarios}
\label{sec:dynamics}

In the discussion above we have made the assumption that the chemical timescales 
are shorter than the dynamical timescale so that chemical equilibrium always holds. 
Whilst this is true for pre-stellar cores in their quasi-static phase of dynamical
evolution, as discussed in RKC24, it may not always be the case; 
for example in the phase of more rapid collapse prior to star formation.
In this section we test whether or not the chemistry can keep up with the dynamics 
in rapidly evolving systems and whether the assumption of chemical equilibrium
still holds. 
To do this, we consider the extreme situation of matter that is undergoing 
free-fall collapse. In the absence of shocks or supersonic flows this equates to
the shortest possible dynamical timescale in normal circumstances.

As before, we have used a model that is based on the {\sc STARCHEM} code, with 
the same chemistry (and including reactive desorption)
but here the chemistry is followed for a single point that is first of all 
evolved in static conditions at n=10$^3$cm$^{-3}$ for 1Myr, and then undergoes 
homologous spherical free-fall collapse to a density of $10^7$cm$^{-3}$. We also 
include an arbitary dynamical retardation factor, as defined in equations 
\ref{eqn:ffall} and \ref{eqn:ffallt}; $B$=0.2, 0.5 or 1, with $B=1$
corresponding to uninhibited free-fall.

In this model, for the sake of simplicity, we keep the temperature fixed;
$T_{\rm gas}=T_{\rm dust}=10$K, and assume that the extinction follows a simple 
power law,
\begin{equation}
{\rm  A_v= 4\left( \frac{n}{2\times 10^4 cm^{-3}}\right)^{0.55}, }
\end{equation}
which gives a good fit in the range ($10^2\leq n \leq 10^6$cm$^{-3}$) to the extinction 
profile at the final timepoint (i.e. near the end of the period of quasi-static
contraction) in the L1544 model.

The rate of change of the density in spherical free-fall collapse is given 
by \citep[e.g.][]{Spitz78}:
\begin{equation}
\frac{dn}{dt} = \left( 24\pi G\rho_0\right)^{1/2} \left( \frac{n}{n_0}\right)^{1/3}n
\left[ \left( \frac{n}{n_0}\right)^{1/3} -1 \right]^{1/2}B
\label{eqn:ffall}
\end{equation}
For a cloud of uniform density that is undergoing free-fall collapse, the 
density remains independent of position within the cloud at all times.
The collapse timescale is
\begin{equation}
\tau_{\rm ff} = \frac{1}{B}\sqrt{\frac{3\pi}{32G\rho_0}} = \frac{5.15\times 
10^7}{B}\left( \mu n_0 \right)^{-1/2}~{\rm years},
\label{eqn:ffallt}
\end{equation}
where $G$ is the gravitational constant, $n_0$ and $\rho_0$ are the initial 
hydrogen nucleon and mass densities, respectively, with
$\rho_0=\mu m_{\rm H}n_0$, and $\mu \sim 1+4X_{\rm He} \sim 1.4$ 
is the mean particle mass per hydrogen nucleon. For the parameters in our model
the collapse timescale is evaluated to be $\sim (1.35/B)$~Myr.

The limiting chemical timescale is that for the freeze-out reactions.
From the formulae in Appendix~\ref{sec:fodes} for oxygen atoms this is
\begin{equation}
{\rm \tau_{freeze} \sim \frac{1.1\times 10^{-3}}{\sigma_{A,0}T_{gas}^{1/2}n} ~s.}
\end{equation}
With $T_{\rm gas}=10$K and the value of $\sigma_{\rm A,0}$ given in 
Table~\ref{tab:params} the ratio of the timescales is
\begin{equation}
{\rm \frac{\tau_{freeze}}{\tau_{ff}} = 1.34\left( 
\frac{10^3 cm^{-3}}{n}\right)^{1/2}. }
\end{equation}
This ratio is less than one for densities 
$\gtappeq 1.8\times 10^3$cm$^{-3}$ and becomes smaller as free-fall 
progresses. The freeze-out efficiency will be lower if the grain surface 
area is reduced due to the coagulation of grains. However, this may not 
be so significant if the highly irregular morphology of the dust grains 
and the geometric growth of grains due to the accumulation of thick ice 
mantles are taken into account. 
We can therefore conclude that, provided the freeze-out is not 
inhibited by desorption processes (at low extinctions), the approximation 
of chemical quasi-equilibrium is likely to be reasonably accurate in most 
high density regions.
%
\begin{figure*}
\includegraphics[scale=0.7]{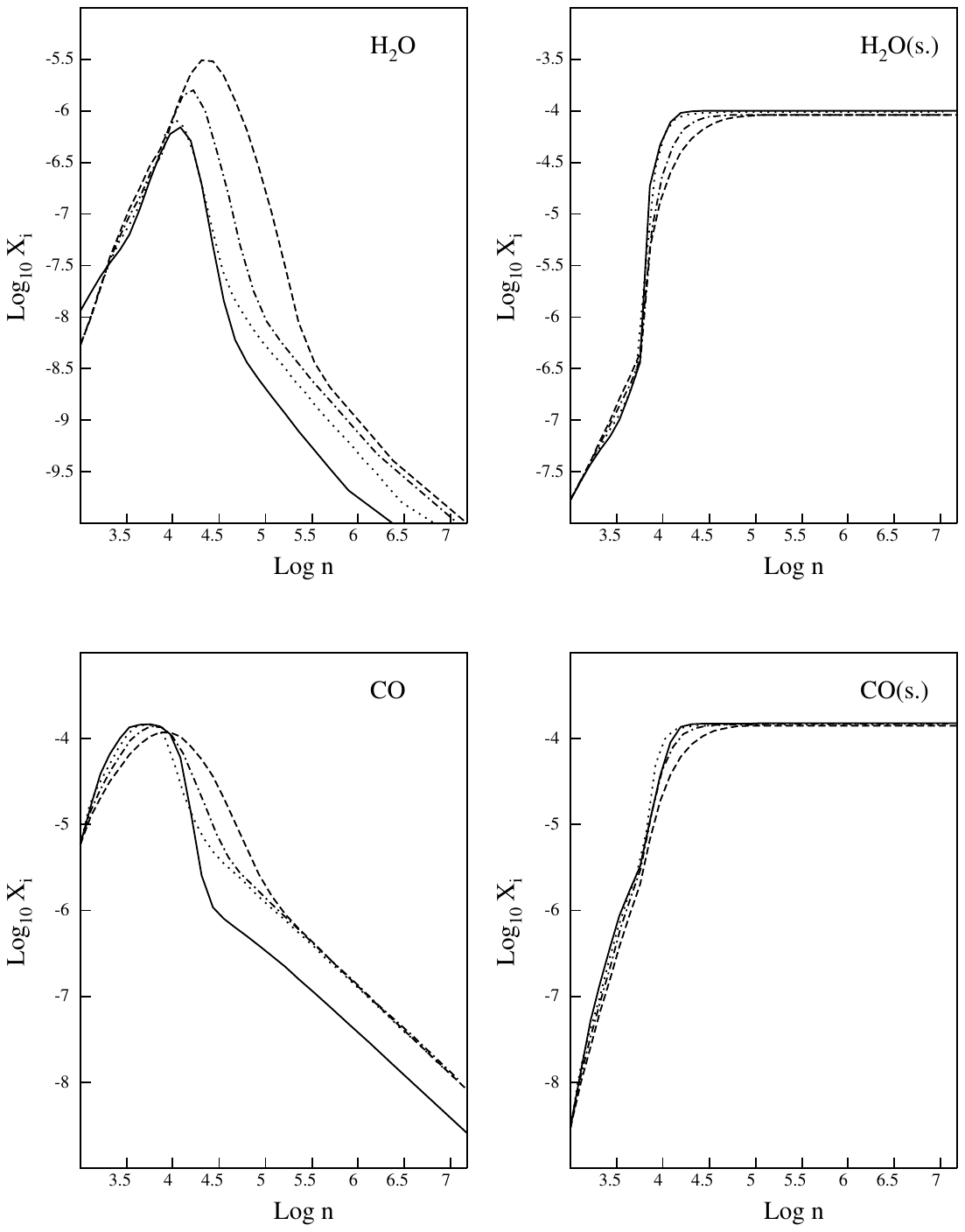} 
\caption{Comparison to free-fall collapse models.
The solid lines show the equilibrium abundances obtained from the 
simplified three-phase network model (with reactive desorption included), 
whilst the dashed, dot-dashed and dotted lines give the results from the full 
model of the time-dependent chemistry ({\sc STARCHEM}) for free-fall 
collapse with values of the collapse retardation factor, 
$B$=1.0, 0.5 and 0.2, respectively. 
$B$=1.0 corresponds to uninhibited free-fall collapse.}
\label{fig:freefall}
\end{figure*}

The results are shown in Figure~\ref{fig:freefall}
which compares the equilibrium abundances obtained from our analytical 
model with those obtained from the {\sc STARCHEM} model of the time-dependent
chemistry. 
Even in the case of free-fall collapse, the simplified network
detailed balance solutions give remarkably good fits to the abundance profiles
obtained from the full model, particularly for the ice components.
Indeed the fit is typically good to a factor of $\sim 2\times$ for the models with 
$B$=0.5 and 0.2 and even for $B$=1.0 there is a good match, except for the period 
when the gas passes through the density range $\sim 10^4-10^5$cm$^{-3}$.
The slight mismatch in the CO and H$_2$O abundance profiles at higher densities is a
consequence of the assumptions about the compositional structure of the ice, rather 
than dynamical effects.

\section{Limitations and assumptions}
\label{sec:limits}

In applications of the simple approximations for calculating the abundances 
of CO, H$_2$O and related chemical species that we describe in this study we must
be aware of the various simplifications, assumptions and limitations of the 
model.
Some of these are generic, in that they apply to most, if not all, astrochemical
models that include gas-grain interactions, whilst others are specific to our model.

In the former category we must include;
\begin{itemize}

\item[(i)] the assumption that the conversion of C$\to$CO is near complete {\em ab initio.}
The conventional approach is to assume that efficient CO formation has already 
occurred when the nascent core was more diffuse and freeze out inhibited (as we
have assumed in our models). But, the formation of CO is relatively slow and 
this may not always be the case.
As explained in RKC24 this leads to competition between 
\begin{equation}
{\rm C ~or~ C^+ + H_2 ~or~ OH ...\to CO  }
\end{equation}
\begin{equation}
{\rm CO + grain \to CO_{\rm (s.)} ...\to CH_3OH ~etc.}
\end{equation}
and
\begin{equation}
{\rm C ~or~ C^+ + grain \to CH_4(s). }
\end{equation}
So, if freeze-out is possible, a significant abundance of CH$_4$ ice may build 
up if there is incomplete conversion of C to CO in the pre-contraction phase.
\item[(ii)] Whilst the dust surface merely acts as a source/sink for the 
gas-phase CO, in the case of the oxygen chemistry, it plays an active 
role in the formation of H$_2$O. 
We assume that the conversion of O and OH to H$_2$O is fast and efficient.
For this to be true, (a) there must be a population of mobile hydrogen atoms on 
the grains and (b) the timescale for diffusion and reaction must be short relative
to the residence timescale for any adsorbed atom/radical. This is generally
true, but may not be the case if there is rapid (i.e. thermal) desorption.
Our solution is primarily adapted for low temperature environments 
($T_{\rm d}\ltappeq 15$K), but it can be applied to warmer regions. However, 
the chemisorption of oxygen atoms is precluded due to the low binding energy
($E_{\rm b}$) when the the dust temperature 
$T_{\rm d}\gtappeq 15$K \citep[if $E_{\rm b}/k=800$K,][]{Schm14} or 
$T_{\rm d}\gtappeq 32$K \citep[if $E_{\rm b}/k=1660$K,][]{He15}. 
In these circumstances the chemical reation network would need to be modified
accordingly.
\item[(iii)] At the microscopic level, the CO and H$_2$O ices are treated as being 
spatially distinct (in the partial monolayer regime) and are assumed to
accumulate in concentric layers, rather than stacking unevenly.
\item[(iv)] We represent the dust grain size distribution by spherical grains 
of a single size, and we do not consider the chemical effects of irregular 
grain morphologies or the aggregation of grains 
\citep[e.g. as discussed in][]{Sip20,Sil21}.
\end{itemize}

The simplifications that are specific to our model include;
\begin{itemize}
\item[(v)] H$_2$O and CO are assumed to be the dominant components of the ice
mantles, and we do not take account of the compositional structure of the ices
so that the relative abundances in the surface layer(s) are taken to the same 
as the bulk composition of the ice.
\item[(vi)] We do not include any allowance for the surface chemistry that 
results in the formation of CO$_2$, CH$_3$OH and other complex organics. 
These species have higher surface binding energies and so are more 
likely to be retained in the ice mantles, especially in those situations 
where thermal desorption is important.
This may be an over-simplification in some circumstances where the relative
abundances of these species are known to be high.
However, the conversion efficiency of CO in surface reactions is $<10\%$, 
so this will not have a significant effect on our results. This has been 
verified by comparison runs of {\sc STARCHEM} which include the surface 
chemistry reactions.
\item[(vii)] We have not included the effects of H$_2$ formation-driven 
desorption \citep{WRW94,Pant21} although, 
as described in Appendix~\ref{sec:fodes}, it would be fairly easy to do this.
\item[(viii)] 
We assume that the fractional abundance of electrons is the same as
that of C$^+$, which is a good approximation at $A_{\rm v}\ltappeq 1.7$
(see Figure~\ref{fig:Xe} in Appendix~\ref{sec:ionize}).
This is also validated by the fact that the abundances of C and C$^+$ in the
simple model are well-matched by {\sc STARCHEM}. In regions where most of the
carbon is in the form of CO, there are additional contributions to the
ionization, in particular the cosmic-ray ionization of H$_2$ and He,
not considered for this study. As a consequence, the fractional ionization in 
the dense gas is highly underestimated (see Figure~\ref{fig:Xe}). 
However, in these circumstances the abundances of C and C+ are insignificantly 
low so that these contributions do not affect the validity of our model.

\end{itemize}

Finally, we should remember that {\sc STARCHEM} includes many simplifications, 
and the three-phase paradigm is itself an approximation, that may not be valid 
if the dust grains have complex and irregular morphologies. 


\section{Conclusions}
\label{sec:summary}

We have re-assessed the use of highly simplified chemical networks for the 
rapid calculation of the abundances of gas-phase and solid-state CO and 
H$_2$O, as well as OH and the important atomic coolants C, C$^+$ and O.

We have presented a modified and updated form for the 
detailed balance solutions, obtained with the assumption of chemical
equilibrium.
This includes a careful re-assessment of the rates for the ice 
desorption proceses in the `three-phase' paradigm. This allows the 
formulation of a set of analytical 
equations that have simple linear dependences on, and can therefore be 
solved for the abundances of 8 chemical species;
C, C$^+$, CO, CO$_{\rm (s.)}$, O, OH, H$_2$O and H$_2$O$_{\rm (s.)}$.

The solution compensates for the non-linearities of the desorption rates in 
regions where there is more than one ice monolayer on the dust grains, takes 
account of geometrical grain growth and the depletion of oxygen into CO.
We have also provided an alternative solution that includes additional terms 
to allow for the reactive desorption of OH and H$_2$O.

In comparisons with fully descriptive chemical models of starless cores we 
find that, in all cases, the model can now well-reproduce the results obtained 
from comprehensive three-phase chemical models with great accuracy
and speed. Thus, using typical current computational resources,
a typical full time-dependent calculation with STARCHEM takes 10-30 minutes,
whilst our analytical solution is obtained in $<$0.1s.
Excellent fits are obtained over the full range of densities
($250\ltappeq n\ltappeq 10^9$cm$^{-3}$) and extinction 
($0.3\ltappeq A_{\rm v}\ltappeq 200$) that we model. 
Apart from a slight mismatch with the OH abundance profiles, the results are
accurate to within a factor of 2 at all positions, which is typically 
less than the uncertainties in the rate coefficients and desorption yields.
This result confirms the findings of RKC24 that the chemistry evolves in 
quasi-equilibrium in pre-stellar cores, and is completely dominated by 
gas-grain processes.

In these comparisons it is important to note that all detailed astrochemical
models, such as {\sc STARCHEM}, are themselves only approximations of reality.
They are, however, fair representations of our current understanding of 
(three-phase) gas-grain interactions.
The fact that our simplified network analytical solution accurately matches the 
results obtained from {\sc STARCHEM} verifies the assumptions and approximations 
that we have made in its formulation.

The solution that we have discussed assumes chemical equilibrium, and has been 
applied to a slowly evolving dynamical systems.
Obviously, the formulation would have limited applicability in situations with
extreme time-variability, such as jets or shocked gas, and does not consider
processes such as turbulent transport on lengthscales that are comparable, or
greater, than those over which significant physical changes occur. 
However, we find that it is also fairly accurate in modelling abundances in
conditions evolving as fast as free-fall collapse.

The main limitations of the model are that for the high density (ice rich) 
regions it assumes that the ice mantles are
predominantly composed of CO and H$_2$O - which may not be applicable if there is
incomplete conversion of C to CO in the low density gas and hydrogenation of
CO into CH$_3$OH in higher density regions, and it assumes that 
the surface hydrogenation of O and OH to H$_2$O is efficient - which is probably
incorrect in regions of low extinction where the dust grains are warm. 

Subject to the caveats mentioned above, we therefore commend this solution
for general application in astrochemical models.

\section*{Data availability}

The data underlying this study are openly available from the published
papers that are cited in the article.
The data generated in support of this research are partly available
in the article, and will be shared on reasonable request to the corresponding
author.

\appendix
\section{Reaction rates}
\label{sec:fodes}

The values of the rate coefficients (in cm$^3$s$^{-1}$) that we adopt in this 
study are:\\
$k_1 = 3.1\times 10^{-13} (T_{\rm g}/300)^{2.7} e^{-3150/T_{\rm g}}$,\\
$k_2 = 2.05\times 10^{-12} (T_{\rm g}/300)^{1.52} e^{-1736/T_{\rm g}}$,\\
$k_3 = 4.67\times 10^{-12} (T_{\rm g}/300)^{-0.6}$,\\
$k_4 = 1.0\times 10^{-10}$,\\
$k_5 = 7.7\times 10^{-10} (T_{\rm g}/300)^{-0.5}$,\\
where $T_{\rm g}$ is the gas kinetic temperature. 

For the photreactions, the reaction data is given by:\\
$P_{\rm OH} = [3.5\times 10^{-10}, -1.7, 254.5]$,\\
$P_{\rm H_2O} = [5.9\times 10^{-10}, -1.7, 485.5]$,\\
$P_{\rm C} = [3.0\times 10^{-10}, -3.0, 255.0]$,\\
$P_{\rm CO} = [2.0\times 10^{-10}, -2.5, 105.0]$,\\
and the rate coefficents (in s$^{-1}$) for $[\alpha, \beta, \gamma]$ are given by
$ k_{\rm i} = \alpha e^{-\beta A_{\rm v}} + \frac{\gamma}{1-\omega}\zeta $,
where $A_{\rm v}$ is the extinction, $\omega$ is the mean grain albedo (taken to be
0.5) and $\zeta$ is the cosmic ray ionization rate.
There are additional modifications to these rates to make allowance for the 
photon doninated region \citep[as described in][Appendix B2]{RKC24}.  

For the freeze-out reactions, following \citet{RHMW92} we can write the 
rate (cm$^{-3}$s$^{-1}$) for a (neutral) species $i$ as
\begin{equation}
\dot{n_{\rm i}} = \dot{X_{\rm i}}n = \sigma_{\rm gr}
\left(\frac{8k_{\rm B}T_{\rm g}}{\pi m_{\rm i}} \right)^{1/2} n_{\rm i}n_{\rm g},
\end{equation}
where $\sigma_{\rm gr}$ is the mean grain cross-section for a given grain
size distribution ($=\pi\langle a^2\rangle$) and $\langle a^2\rangle$ is 
the mean of the square of the grain radius.
The term in brackets is the thermal velocity of the atom/molecule and
we have assumed that all impacting atoms/molecules stick to the grains.
$k_{\rm B}$ is the Boltzmann constant and $n_{\rm g}$ is the number density 
of grains ($n_{\rm g}=d_{\rm g}n$, where $d_{\rm g}$ is the dust-to-gas ratio, 
by number).

Assuming that the grains are spherical,
the dust grain surface area per hydrogen nucleon $\sigma_{\rm A} =
4\sigma_{\rm gr}d_{\rm g}$. 
Hence
$ \dot{X_{\rm i}} =  F_{\rm i}X_{\rm i}n$, with
\begin{equation}
F_{\rm i} = \left(\frac{k_{\rm B}}{2\pi m_{\rm H}} \right)^{1/2}
\sigma_{\rm A} \left( \frac{T_{\rm g}}{\mu_{\rm i}}\right)^{1/2},
\end{equation}
where $m_{\rm H}$ is the hydrogen atom mass and $\mu_{\rm i}$ is the mass of 
species $i$ in $amu$.

For the desorption rates, we note that these primarily occur due to the release 
of bound species from the surface of grains and the rates for these processes 
are given by
\begin{equation}
\dot{X_{\rm i}} = D_{\rm des}f_{\rm i},
\end{equation}
where $D_{\rm des}$ is the total desorption rate (s$^{-1}$) which we take to be the 
sum of thermal desorption, photodesorption (direct and cosmic-ray induced),
cosmic-ray heating induced desorption and desorption driven by the enthalpy of
formation of a species on grain surface: 
\begin{equation}
D_{\rm des} = D_{\rm th} + D_{\rm pd} + D_{\rm cr} +D_{\rm H} 
\end{equation}
and $f_{\rm i}$ is the fraction of the grain surface that is covered by 
species $i$. 
The individual components of $R_{\rm des}$ (as discussed in RKC24) are:

Thermal sublimation, for which the rate is given by
\begin{equation}
D_{\rm th} = \nu_{\rm i} e^{-T_{\rm b}/T_{\rm d}} \sigma_{\rm A} N_{\rm s}, 
\end{equation}
where $\nu_{\rm i}$ is the lattice vibrational frequency of the adsorbed 
species:
\begin{equation}
\nu_{\rm i} = \sqrt{\frac{2N_{\rm s}k_{\rm B}T_{\rm b}}{\pi^2 \mu_{\rm i} m_{\rm H}}} 
\sim 10^{12}-10^{13} s^{-1}. 
\end{equation}
$T_{\rm b}$ is the binding temperature of the absorbed species
(related to the adsorption binding energy by $T_{\rm b}=E_{\rm i}/k_B$) 
and $T_{\rm d}$ is the dust temperature.

Photodesorption, with a rate (assuming an isotropic radiation field) given by,
\begin{equation}
D_{\rm pd} = \left[ G_0I_0 e^{-1.8A_{\rm v}} + 
I_{\rm cr} \left(\frac{\zeta}{\zeta_0}\right) \right] Y_{\rm i} \sigma_{\rm A}, 
\end{equation}
where $I_0$ is the interstellar UV photon flux, $G_0$ is the radiation field 
scaling factor, $A_{\rm v}$ is the extinction, 
$I_{\rm cr}$ is the cosmic-ray induced photon flux, 
$(\zeta/\zeta_0)$ is the ratio of the cosmic-ray ionization rate to 
$\zeta_0 = 1.3\times 10^{-17}$s$^{-1}$ (note that, although $I_{\rm cr}$ is 
sometimes specified as a fraction of $I_0$, the two quantities are 
independent) and 
$Y_{\rm i}$ is the photodesorption yield per photon for species $i$.
The photodesorption of H$_2$O can either be non-dissociative (with yield 
$Y_{\rm H_2O}$) or dissociative, resulting in the formation of OH, or O
(with yields of $Y_{\rm OH}$ and $Y_{\rm O}$, respectively).
In the applications discussed in this paper, we follow the practice of 
\citet{Holl09} and assume that $Y_{\rm OH}=2Y_{\rm H_2O}$. 

Desorption driven by the cosmic-ray heating of grains, for which the rates 
are given by
\begin{equation}
D_{\rm cr} = r_{\rm cr,i}\sigma_{\rm A} N_{\rm s}, 
\end{equation}
where $r_{\rm cr,i}$ is the cosmic-ray heating desorption rate (s$^{-1}$)
per molecule for species $i$, which does not necessarily have a linear
dependence on $\zeta$ \citep[][table 9]{Raw22}, and may result in the 
desorption of multiple ice layers.

We have not included H$_2$ formation driven desorption in this model.
As the rate of H$_2$ formation is defined by the gas-phase hydrogen atom 
density, it would be relatively easy to do so on the basis of analytically
determined H$:$H$_2$ ratios. 

\section{The analytical solutions}
\label{sec:analytic}

In this section we give the full analytical solution to the matrix representation
of the H$_2$O chemical network and the equations for the
CO chemistry, as given in Figures~\ref{fig:matrix} and \ref{fig:numatrix}.

To be concise we have labelled the various terms for the rates for the H$_2$O
chemistry, as given in Section~\ref{sec:networks} and Section~\ref{sec:rdes}, 
with the letters $a-m$ according to the table below:

\begin{tabular}{ll}
\hline
$k_1n/2$ & a \\
$k_2n/2$ & b \\
$P_{\rm OH}$ & c \\
$P_{\rm H_2O}$ & d \\
$F_{\rm O}n$ & e \\
$F_{\rm OH}n$ & f \\
$F_{\rm H_2O}n$ & g \\
$D_{\rm H_2O}/(\sigma_{\rm A}N_{\rm s})$ & h \\
$D_{\rm OH}/(\sigma_{\rm A}N_{\rm s})$ & i \\
$D_{\rm O}/(\sigma_{\rm A}N_{\rm s})$ & j \\
\hline
$H_1n$ & k \\
$H_2n$ & l \\
$H_3n$ & m \\
\hline
\end{tabular}
~\\

\noindent
Following the determinant and adjoint method of matrix inversion,
we write sums of the products of the various rate terms:\\
$\alpha =$ jbg + jcd + jcg + jdf + jfg + ceh + cdi + cgi\\
$\beta =$  jad + jag + adh + adi + agi + deh + dei + egi\\
$\gamma =$ jab + abh + abi + afh + beh + bei + ceh + efh\\
$\delta =$ abg + adf + afg + beg + cde + ceg + def + efg\\

\noindent
with modifications, if the effects of reactive desorption are 
to be included:

\noindent
$\alpha = \alpha -$ jmd\\
$\beta = \beta +$ jkd + jkg + jld\\
$\gamma = \gamma +$ jkm + jkb + jlb + jlc + jlf + jma -kch + lci + mai + mei\\
$\delta = \delta -$ kcd - kcg - lcd - mad - mde\\

\noindent
The solutions for the equilibrium fractional abundances are then:\\

\noindent
X(O) = $X_{\rm O,total}\alpha$/($\alpha + \beta + \gamma + \delta$)\\
X(OH) = $X_{\rm O,total}\beta$/($\alpha + \beta + \gamma + \delta$)\\
X(H$_2$O) = $X_{\rm O,total}\gamma$/($\alpha + \beta + \gamma + \delta$)\\
X(H$_2$O$_{(\rm s.)}$) = $X_{\rm O,total}\delta$/($\alpha + \beta + \gamma + \delta$)\\

\noindent
Note that the solution is considerably simplified if one wishes to 
omit the two high temperature gas-phase reactions, in which case $a=b=0$.

\noindent
Similarly, for the CO chemistry, we abbreviate the rates for the processes as 
given in Section~\ref{sec:networks} according to the table below:

\begin{tabular}{ll}
\hline
$P_{\rm C}$ & n \\
$k_3nX_{\rm C,total}$ & p \\
$k_4nX_{\rm OH}$  & q \\
$k_5nX_{\rm OH}$  & r \\
$P_{\rm CO}$ & s \\
$F_{\rm CO}n$ & t \\
$D_{\rm CO}/(\sigma_{\rm A}N_{\rm s})$ & u \\
\hline
\end{tabular}
~\\

\noindent
Writing:\\
v = su/(su +qu +qt)\\
w = (su+ ru + rt)/(su + qu + qt)

\noindent
then the solutions for the equilibrium fractional abundances are:\\
~\\
$X(C^+)= X_{\rm C,total}\left[ \sqrt{(r+nw)^2 + 4pnv} - (r+nw)\right] /2p $ \\
$X(C) = {\rm v}X_{\rm C,total}-qX(C^+)$ \\
$X(CO) = \left(\frac{u}{t+u}\right) \left[ X_{\rm C,total} - X(C^+) - X(C)\right] $ \\
$X(CO_{\rm s.}) = (t/u)X(CO)$ 

\section{Including the effects of grain growth}
\label{sec:growth}

In this paper we have followed previous studies and have approximated the dust 
grain size distribution by grains of a single size with properties of the mean 
of the distribution. 
In the discussions above we have made the simplification that the size of the
grains is fixed and not time-dependent.
Neglecting processes such as agglomeration and sputtering 
this is a fair approximation for large grains ($a\gtappeq 0.1\mu$m) but for very 
small grains ($a\ltappeq 0.01\mu$m) the thickness of the accumulated ice mantles 
may be comparable to, or larger, than the radius of the bare dust grain. 
In these circumstances the effect of the larger grain surface area may be 
significant in the determination of the freeze-out and desorption rates.

If we make the simplification that the grain size distribution can be represented 
by a single size, equal to the rms grain radius for the distribution, then 
this condition corresponds to ($X_{\rm ice}\Delta r/N_{\rm s}\sigma_{\rm A})\gtappeq a_0$ and will be applicable for small initial grain sizes (e.g. $a_0\sim 0.01\mu$m).

The volume occupied by each ice molecule is $\sim (\Delta r/N_{\rm s})$. 
If $a$ and $a_0$ are the radii of the grains with/without the ice mantle,
respectively, then the volume of the ice mantle is $\frac{4\pi}{3}(a^3 -a_0^3)$, 
so that the number of ice molecules per grain is 
$\frac{4\pi}{3}\frac{N_{\rm s}}{\Delta r}(a^3 -a_0^3)$.

If $n_{\rm gr}$ is the number density of dust grains, then 
the number of ice molecules per unit volume is
$X_{\rm ice}.n_{\rm H} = 
\frac{4\pi}{3}\frac{N_{\rm s}}{\Delta r}(a^3 -a_0^3)n_{\rm gr}$,
where $n_{\rm H}$ is the hydrogen nucleon density.

The surface area of the bare grains, per hydrogen nucleon, is
$\sigma_{\rm A,0} = 4\pi a_0^2 n_{\rm gr}/n_{\rm H}$.

Hence the relationship between the grain size ($a$) and the ice abundance 
($X_{\rm ice}$) is
\begin{equation}
a^3 = \left( \frac{3a_0^2\Delta r}{\sigma_{\rm A,0} N_{\rm s}}\right) 
X_{\rm ice} + a_0^3 \mbox{~~cm$^3$,}
\label{eqn:growth}
\end{equation}
where we have assumed that each ice layer has uniform thickness ($\Delta r$, 
equal to the mean inter-molecular spacing) and 
the same density of binding sites ($N_{\rm s}$).
The dust surface area will scale accordingly:
\begin{equation}
\sigma_{\rm A} = \sigma_{\rm A,0} \left( \frac{a}{a_0} \right)^2 \mbox{~cm$^2$}.
\label{eqn:area}
\end{equation}
For small grains ($a\sim 0.01\mu$m) this scaling factor is $\sim 5$ and should 
be included in the calculation of $R_{\rm scale}$.

For large grains ($a\gtappeq 0.1\mu$m) the geometric growth will not be so 
significant and it is satisfactory to use the approximation that 
$\sigma_{\rm A} = \sigma_{\rm A,0}$.

\section{Fractional ionization}
\label{sec:ionize}

\begin{figure}
\includegraphics[width=0.85\columnwidth,trim=0.5in 5.4in 4.1in 1in,clip=true]
{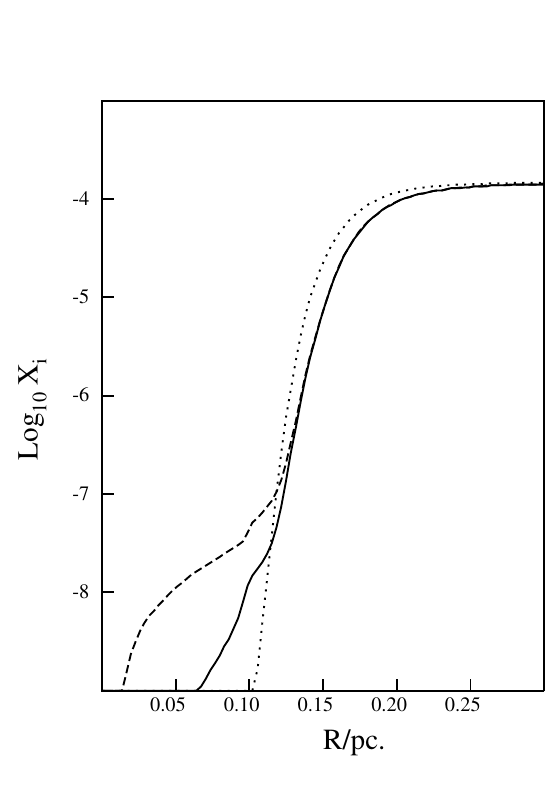}
\caption{Abundance profiles for C$^+$ and electrons: from the full three-phase 
{\sc STARCHEM} chemical model (C$^+$: solid line, e$^-$: dashed line), 
and C$^+$ from the algebraic solution described in this study (dotted lines).}
\label{fig:Xe}
\end{figure}

In figure~\ref{fig:Xe} we compare the abundance profiles (using the physical 
model described in section~\ref{sec:compare}) for C$^+$ and e$^-$
from the {\sc STARCHEM} model with the profile for C$+$ from the algebraic 
solution.
This figure shows that the approximation that $X_{\rm C^+} = X_{\rm e^-}$ 
is valid over a range in the abundance of C$^+$ of three orders in magnitude; 
10$^{-4}$ to 10$^{-7}$, in both the atomic and molecular phases. In the 
central regions where the external UV has been sufficiently attenuated 
($A_{\rm v}\gtappeq 1.7$), $X_{\rm C^+}$ falls below 10$^{-7}$. 
Here, the ionization of H$_2$ and He by cosmic rays becomes a more important contributor to the electron abundance. In addition, the dominant production
mechanism of C$^+$ shifts from photoionization to charge exchange of C 
with He$^+$.

\label{lastpage}


\begin{thebibliography}{}

\bibitem[Aikawa et al.(2001)]{Aik01}
Aikawa Y., et al., 2001, ApJ, 552, 639.

\bibitem[Aikawa et al.(2005)]{AHRC05}
Aikawa Y., Herbst E., Roberts H., Caselli P., 2005, ApJ, 620, 330

\bibitem[Broderick et al.(2007)]{Brod07}
Broderick A.E., Keto E., Lada C.J., Narayan R., 2007, ApJ, 671, 1832

\bibitem[Caselli et al.(2010)]{Cas10} 
Caselli P., et al., 2010, A\&A, 521, L29

\bibitem[Caselli et al.(2012)]{Cas12} 
Caselli P., Keto E., Bergin E.A. et al., 2012, ApJL, 759, L37

\bibitem[Furuya et al.(2017)]{FDV17}
Furuya K., Drozdovskaya M., Visser R., et al., 2017, A\&A 599, A40

\bibitem[Garrod(2013)]{Garr13}
Garrod R.T., 2013, ApJ, 765, 60

\bibitem[Glover \& Clark(2012)]{GC12}
Glover S.C.O., Clark P.C., 2012, MNRAS, 421, 116

\bibitem[Hasegawa \& Herbst(1993)]{HH93}
Hasegawa T.I., Herbst E., 1993, MNRAS, 263, 589

\bibitem[He et al.(2015)]{He15}
He J., Shi J., Hopkins H., Vidali G., Kaufman M.J., 2015, ApJ, 801, 120

\bibitem[Hollenbach et al.(2009)]{Holl09}
Hollenbach D., Kaufman M.J., Bergin E.A., Melnick G.J., 2009, ApJ, 690, 1497

\bibitem[Keto \& Caselli(2008)]{KC08}
Keto E., Caselli P., 2008, ApJ, 638, 238 

\bibitem[Keto \& Caselli(2010)]{KC10}
Keto E., Caselli P., 2010, MNRAS, 402, 1625 

\bibitem[Keto, Rawlings \& Caselli(2014)]{KRC14}
Keto E., Rawlings J.M.C., Caselli P., 2014, MNRAS, 440, 2616 [KRC14]

\bibitem[Keto, Caselli \& Rawlings(2015)]{KCR15}
Keto E., Caselli P., Rawlings J.M.C., 2015, MNRAS, 446, 3731

\bibitem[Minissale et al.(2016)]{Min16}
Minissale M., Dulieu F., Cazaux S., Hocuk S., 2016, A\&A, 585, A24

\bibitem[Pantaleone et al.(2021)]{Pant21}
Pantaleone S., Enrique-Romero J., Ceccarelli C., Ferrero S., Balucani N., 
Rimila A., Ugliengo P., 2021, ApJ, 917, 49

\bibitem[Rawlings et al.(1992)]{RHMW92}
Rawlings J.M.C., Hartquist T.W., Menten K., Williams D.A., 1992,
MNRAS, 255, 471

\bibitem[Rawlings (2022)]{Raw22} 
Rawlings J.M.C., 2022, MNRAS, 517, 3804

\bibitem[Rawlings, Keto \& Caselli(2024)]{RKC24}
Rawlings J.M.C., Keto E., Caselli P., 2024, MNRAS, 530, 3986 [RKC24]

\bibitem[Riedel et al.(2023)]{Ried23}
Riedel W., Sipil\"{a} O., Redaelli E., Caselli P., Vasyunin A.I., 
Dulieu F., Watanabe N., 2023, A\&A, 680, 87

\bibitem[Roberts et al.(2007)]{RRVW07}
Roberts J.F., Rawlings J.M.C., Viti S., Williams D.A., 2007, MNRAS, 382, 733

\bibitem[Ruaud, Wakelam \& Hersant(2016)]{RWH16}
Ruaud M., Wakelam V., Hersant F., 2016, MNRAS, 459, 3756

\bibitem[Schmalzl et al.(2014)]{Schm14}
Schmalzl M., Visser R., Walsh C., Albertsson T., Kristensen L.E., 
Mottram J.C., van Dishoeck E.F., 2014, A\&A, 572, A81

\bibitem[Shingledecker et al.(2018)]{Shin18}
Shingledecker C.N., Tennis J., Le Gal R., Herbst E., 2018, 
ApJ, 861, 20  

\bibitem[Silsbee, Caselli \& Ivlev(2021)]{Sil21}
Silsbee K., Caselli P., Ivlev A.V., 2021,
MNRAS, 507, 6205

\bibitem[Sipil\"{a}, Zhao \& Caselli(2020)]{Sip20}
Sipil\"{a} O., Zhao B., Caselli P., 2020,
A\&A, 640, 94

\bibitem[Spitzer (1978)]{Spitz78}
Spitzer L., 1978, in `Physical Processes in the Interstellar Medium' 
(John Wiley, New York)

\bibitem[Taquet, Ceccarelli \& Kahane(2012)]{TCK12}
Taquet V., Ceccarelli C., Kahane C., 2012, A\&A, 538, A42

\bibitem[van Dishoeck et al.(2013)]{vD13}
van Dishoeck E.F., Herbst E., Neufeld D.A., 2013, 
Chemical Reviews, 113, 9043

\bibitem[van Dishoeck et al.(2021)]{vD21}
van Dishoeck E.F., Kristensen L.E., Mottram J.C., et al., 2021, 
A\&A, 648, 24

\bibitem[Willacy et al.(1994)]{WRW94} 
Willacy K., Rawlings J.M.C., Williams D.A., 1994, MNRAS, 269, 921

\end{thebibliography}
\end{document}